\documentclass[9pt, twocolumn]{article}
\usepackage{graphicx}
\usepackage{caption}
\usepackage{amsmath,amssymb}
\usepackage{authblk}
\usepackage[letterpaper, total={7.5in, 9.5in}]{geometry}
\title{Active-cavity photonic molecule optical data wavelength converter for silicon photonics platforms}

\author[1,*]{Hayk Gevorgyan}
\author[1,2]{Anatol Khilo}
\author[1,$\dagger$]{Milo\v{s} A. Popovi\'{c}}

\affil[1]{Department of Electrical and Computer Engineering, Boston University, Boston, MA 02215, USA}
\affil[2]{Currently with Ayar Labs Inc., Emeryville, CA 94608, USA}
\affil[ ~]{Corresponding authors: $^*$hayk@bu.edu, $^\dagger$mpopovic@bu.edu}

\begin{document}
\twocolumn[
\begin{@twocolumnfalse}
\maketitle
\begin{abstract}
\textbf{We demonstrate an optical data wavelength converter based on an electrically driven photonic molecule structure comprising two coupled active silicon microring resonators. The converter, supplied by a continuous-wave (CW) microwave drive signal equal in frequency to the desired wavelength shift, replicates an input optical signal at a new wavelength. The optical coupled-cavity system matches supermode resonances to the input and wavelength-converted optical waves maximizing the conversion efficiency. Two device designs that perform wavelength up- and down-conversion by 0.19~nm (24~GHz) with -13~dB conversion efficiency and 6~GHz bandwidth; and by 0.45~nm (56~GHz) with -18~dB efficiency and 5~GHz bandwidth are demonstrated. A 4~Gbps non-return-to-zero (NRZ) optical data stream is shifted in wavelength and successfully recovered. This architecture accepts CMOS-level RF drive voltages and can be integrated in monolithic CMOS electronic-photonic platforms with a simple signal source circuit as part of a self-contained subsystem on chip that generates and carries out the wavelength conversion, requiring no high-frequency (optical or electrical) and high-power external pump input to the chip. This type of device may become a new standard element in the component libraries of silicon and CMOS photonics processes.}
\bigskip
\end{abstract}
\end{@twocolumnfalse}
]

\section{Introduction}

Conversion of optical data signals between the wavelength channels of wavelength division multiplexed (WDM) optical fiber communication systems has been shown to enable efficient use of network resources and increase capacity \cite{Yoo1996, Elmirghani2000}. Short-reach, high-bandwidth optical WDM data communication systems based on integrated silicon and CMOS photonics technologies can potentially benefit from the wavelength conversion functionality \cite{Stojanovic2018,Atabaki2018,Settaluri2015}, which is not readily available in these technologies with high efficiency and low energy cost.

All-optical wavelength conversion based on a number of physical processes has been studied as an all-optical alternative to optical-electrical-optical conversion. Optical gating schemes, using cross-gain or -phase modulation between probe and pump signals, have been demonstrated in semiconductor optical amplifiers (SOA) \cite{Yoo1996, Durhuus1994}. Wavelength conversion in periodically poled lithium niobate waveguides has been achieved by cascading second harmonic and difference frequency generation processes \cite{Chou1999, Umeki2010}. Techniques based on four-wave-mixing (FWM) have been demonstrated in SOAs and in passive devices, such as optical fibers \cite{Hu2010}, silicon waveguides \cite{Rong2006, Yamada2006}, resonators built on gallium arsenide \cite{Absil2000}, silica glass \cite{Fasquazi2010}, and silicon \cite{Morichetti2011, Ong2014}. Using the combination of two-photon absorption and free carrier dispersion effects for all-optical wavelength conversion has been demonstrated in silicon microring resonators \cite{Xu2005a, Li2008}. In these devices, the data pattern of an input signal coupled to one resonant mode of a silicon cavity modulates all cavity resonances, imprinting the data pattern on a new CW carrier wavelength that is coupled to a different resonant mode. All-optical approaches for wavelength conversion eliminate the need for optical-electrical-optical conversion, but require an additional source of optical energy: a high-power optical pump, or a new carrier at a target wavelength. In many cases the required pump power can reach tens to hundreds of milliwatts [ref.]. In \cite{wade2014}, we proposed an integrated coupled-cavity wavelength converter architecture that does not require an optical source to be provided and relies only on high-frequency electrical modulation of cavity resonances with low-voltage signals. As such, if they can show compelling performance, they can straightforwardly become a new device offering and functionality in current silicon and CMOS photonics platforms \cite{Pantouvaki2017,Fahrenkopf2019,Stojanovic2018,Atabaki2018}. The proposed device overcomes the efficiency roll off problem of single-cavity modulators \cite{Ehrlichman2018}, imposed by the cavity photon lifetime, and shows a fundamental advantage over them of 15 to 50~dB in conversion efficiency \cite{Gevorgyan2020}. Such active devices with two coupled resonators have also been studied for RF-to-optical conversion \cite{Gevorgyan2020, Gevorgyan2019, Zhang2019}, and used as data modulators \cite{Yu2014}.

\begin{figure}[!b]
\centering
\mbox{\includegraphics[width=\linewidth]{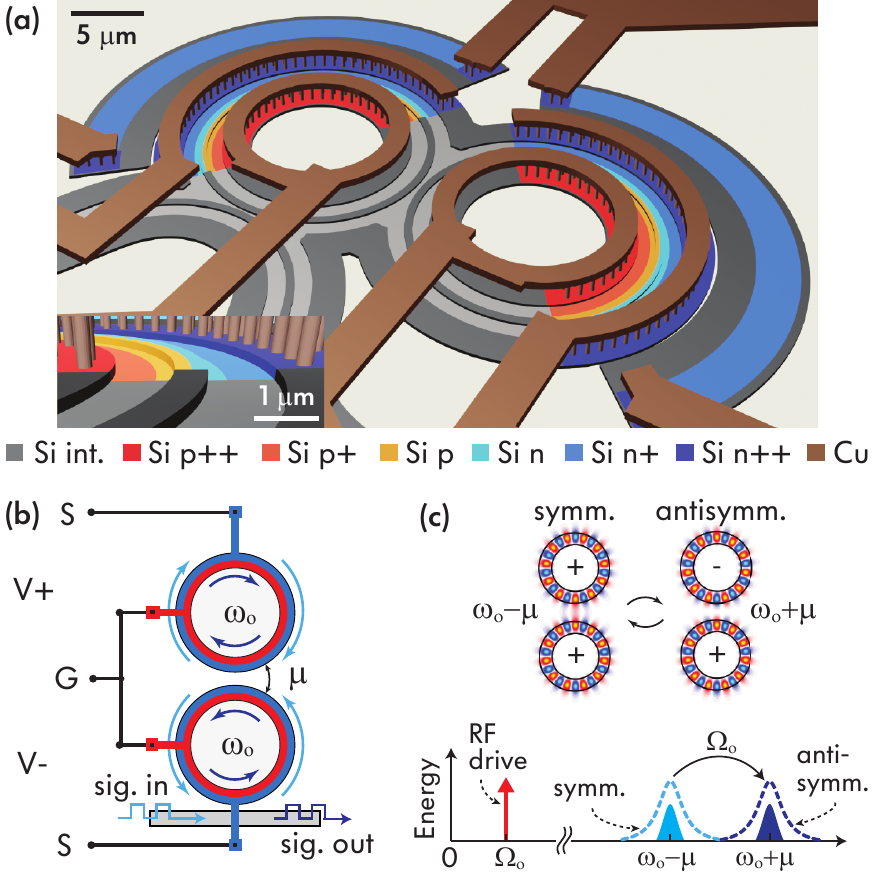}}
\caption{(a) Three-dimensional rendering and (b) schematic representation of the dual-active-cavity wavelength converter. (c) Symmetric and antisymmetric supermodes of the coupled-cavity system (top), which, when coupled due electrical modulation, replicate input signal at a new center wavelength (bottom).}
\label{fig:wc_concept}
\end{figure}

In this work, we report the first demonstration of optical signal wavelength conversion in a dual-active-cavity microring wavelength converter, illustrated in \mbox{Fig.~\ref{fig:wc_concept}(a)}, which is fabricated in a silicon photonics platform. Two device variants that perform wavelength conversion by $\pm$0.19~nm ($\mp$24~GHz) and $\pm$0.45~nm ($\mp$56~GHz) with respective bandwidths of 6 and 4.8~GHz and conversion efficiencies of -13 and -18~dB are experimentally demonstrated. Operation of these devices is validated by shifting the center wavelength of a 4~Gbps NRZ optical data signal with speed limit primarily imposed by the bandwidth of off-chip filtering. The studied converter architecture is also capable of bidirectional operation between two wavelength channels, performing up-conversion in one propagation direction and down-conversion in the opposite direction with equal efficiency, while passing the remaining channels through unaffected. With monolithic CMOS electronic-photonic processes \cite{Stojanovic2018,Atabaki2018}, such devices can be co-integrated with electronic circuits that generate the high-frequency sinusoidal RF drive on chip. Such self-contained on-chip building blocks could perform wavelength conversion without requiring external high frequency and high power auxiliary input (RF or optical) to the chip, only DC power, and could give rise to a new family of silicon photonics process design kit (PDK) components for wavelength-manipulation-based signal processing. With sufficiently improved efficiencies, such devices could provide "beam splitting" functionality for frequency bin qubits for quantum photonic information processing \cite{Lu2018}. 

\section{Operation principles and design}
\label{sec:operation}

The wavelength converter is schematically illustrated in \mbox{Fig.~ \ref{fig:wc_concept}(b)}. Two identical microring resonators are evanescently coupled to each other and to a bus waveguide. The resonators are active and have built-in electro-optic (EO) phase shifters with a common ground and independent signal contacts. 

The system of two coupled resonators supports two resonant supermodes, with symmetric and antisymmetric electromagnetic field distributions in the cavities, as shown in \mbox{Fig.~\ref{fig:wc_concept}(c)}~[top]. Resonance frequencies of the symmetric ($\omega_o-\mu$) and antisymmetric ($\omega_o+\mu$) supermodes are determined by the resonance frequency of the cavities in isolation, $\omega_o$, and the energy coupling rate $\mu$ between them when coupled. Desired resonance frequency splitting $2\mu$ can be achieved by a proper choice of coupling strength (gap spacing) between the rings. With no modulating signal applied, the supermode resonances are uncoupled and no optical energy is transferred from one mode to the other. Modulating individual resonators in push-pull mode with frequency $\Omega_o=2\mu$ creates time-periodic antisymmetric perturbation of refractive index in the resonators, producing an optimal electromagnetic field overlap between the supermodes, which results in energy coupling between them \cite{wade2014, Gevorgyan2020}. This temporal coupling process has a spatial analogue, namely coupling between two non-degenerate modes of a waveguide, phase-matched by a periodic grating (e.g. sidewall corrugation) along the propagation direction \cite{Wang2019}. 

Operation of the wavelength converter is based on the modulation-induced coupling described above. The input optical signal couples into the converter through the bus waveguide and excites one of the two supermode resonances, which is aligned in center frequency with the signal, as shown in \mbox{Figs.~\ref{fig:wc_concept}(b,c)}. The converter, driven by a differential CW electrical signal with frequency $\Omega_o$ equal to the desired conversion frequency, transfers spectral components of the optical signal from the first resonance to the second, effectively converting the wavelength of the input signal. By design, the resonances are separated from each other in frequency by $2\mu=\Omega_o$ [see \mbox{Fig.~\ref{fig:wc_concept}(c)}]\cite{Gevorgyan2020}. The device performs frequency up-conversion (wavelength down-conversion) if the symmetric resonance is tuned to the input signal frequency and frequency down-conversion (wavelength up-conversion) if the antisymmetric resonance is aligned with input signal. The remaining wavelength channels, far from the resonance wavelength, pass through the converter unaffected. The device is also capable of performing simultaneous up- and down-conversion of optical data signals in opposite propagation directions between two wavelength channels, by utilizing a pair of forward and counter propagating supermode resonances. Such functionality is useful for bidirectional optical links where signals are transmitted and received through the same physical medium (e.g. optical fiber, waveguide).

A closed-form formula for the efficiency (conversion power gain) of the converter, derived in \cite{wade2014,Gevorgyan2020}, is given by
\begin{equation}\label{eqn:conveff}
	\left.\begin{aligned}
      &G = \frac{\left(1-\frac{\delta f_{int}}{\delta f_{tot}}\right)^2\left(\frac{\delta f_{pp}}{\delta f_{tot}}\right)^2}{\left(\frac{4\Delta f}{\delta f_{tot}}\right)^2+\left[1 + \left(\frac{\delta f_{pp}}{2\delta f_{tot}}\right)^2-\left(\frac{2\Delta f}{\delta f_{tot}}\right)^2\right]^2},
	\end{aligned}\right.
\end{equation}
where $\delta f_{int}$ and $\delta f_{tot}$ are the intrinsic and total linewidths of the supermode resonances, $\delta f_{pp}$ is the peak-to-peak frequency swing of the cavity resonances due to modulation, and $\Delta f$ is the input signal detuning from a supermode resonance frequency. For given $\delta f_{tot}$ the peak conversion efficiency ($\Delta f=0$) increases with $\delta f_{pp}$ and at $\delta f_{pp}=2\delta f_{tot}$ reaches its maximum, which is limited by the intrinsic linewidth $\delta f_{int}$. The throughput bandwidth of the converter, defined as the full width at half maximum of the conversion efficiency G is given by
\begin{equation}\label{eqn:bandwidth}
	\left.\begin{aligned}
      &BW = \delta f_{tot}\sqrt{\sqrt{2+2\left(\frac{\delta f_{pp}}{2\delta f_{tot}}\right)^4}-1 + \left(\frac{\delta f_{pp}}{2\delta f_{tot}}\right)^2}.
	\end{aligned}\right.
\end{equation}
Under weak ($\delta f_{pp} \to 0$) and optimum ($\delta f_{pp}=2\delta f_{tot}$) modulation conditions the bandwidth as given by Eq.~\ref{eqn:bandwidth} reduces to
\begin{equation}\label{eqn:bandwidth_limits}
	\left.\begin{aligned}
      &BW\big|_{\delta f_{pp} \to 0} = 0.64~\delta f_{tot}, \quad BW\big|_{\delta f_{pp}=2\delta f_{tot}} = 1.41~ \delta f_{tot}.
	\end{aligned}\right.
\end{equation}
Resonance linewidth $\delta f_{tot}$ can be adjusted to produce desired bandwidth by controlling the ring-to-bus coupling strength. Due to a fundamental tradeoff between the bandwidth and conversion efficiency \cite{Gevorgyan2020}, the ring-to-bus power coupling should be designed for maximum conversion efficiency and for a bandwidth wide enough to accommodate input signal spectrum.

The wavelength converter, designed for a foundry silicon photonics process with 220~nm-thick device layer is illustrated in \mbox{Fig.~\ref{fig:wc_concept}(a)}. Two device variants are considered in this work; one designed for frequency shift of 24~GHz, another for 56~GHz. The microring resonators and the bus waveguide are built using rib waveguides with side slab thickness less than 100~nm and a core width of 500~nm. The ring radius and the radius of curvature of the bus waveguide in the ring-to-bus coupling region are 7.5 $\mu$m, measured to the center of the waveguide core. The coupling gap between the ring and bus waveguides is set to 220~nm. The ring-to-ring coupling gap, the only design parameter that is different in the two device variants, is set to 445~nm in the first and 315~nm in the second variant. Electro-optic phase shifters are implemented with lateral-junction p-n diodes built into the core of the ring waveguides. The estimated free carrier concentrations of electrons and holes in the n and p regions of the junction are $3\times10^{17}$ cm$^{-3}$. Electrical vias are placed on 220~nm-thick silicon contact rails, located on either side of the waveguide core, at 1~$\mu$m distance from it [see inset in \mbox{Fig.~\ref{fig:wc_concept}(a)}]. Due to the geometry of the ring phase shifter that has metal contacts on both its inner and outer radii, only half of the circumference of each microring is made active, as shown in \mbox{Fig.~\ref{fig:wc_concept}(a)}. This results in reduced resonance frequency swing $\delta f_{pp}$ and lower conversion efficiency for given drive voltage. The whole cavity could be made active by using more complex phase-shifter geometries with metal contacts only on the inner radius of the microring \cite{watts2014,Stojanovic2018}. Finally, microheaters adjacent to the cavities are designed to enable tuning of their resonances.

\section{Experimental demonstration}
\label{sec:experiment}

\begin{figure}[!b]
\centering
\mbox{\includegraphics[width=\linewidth]{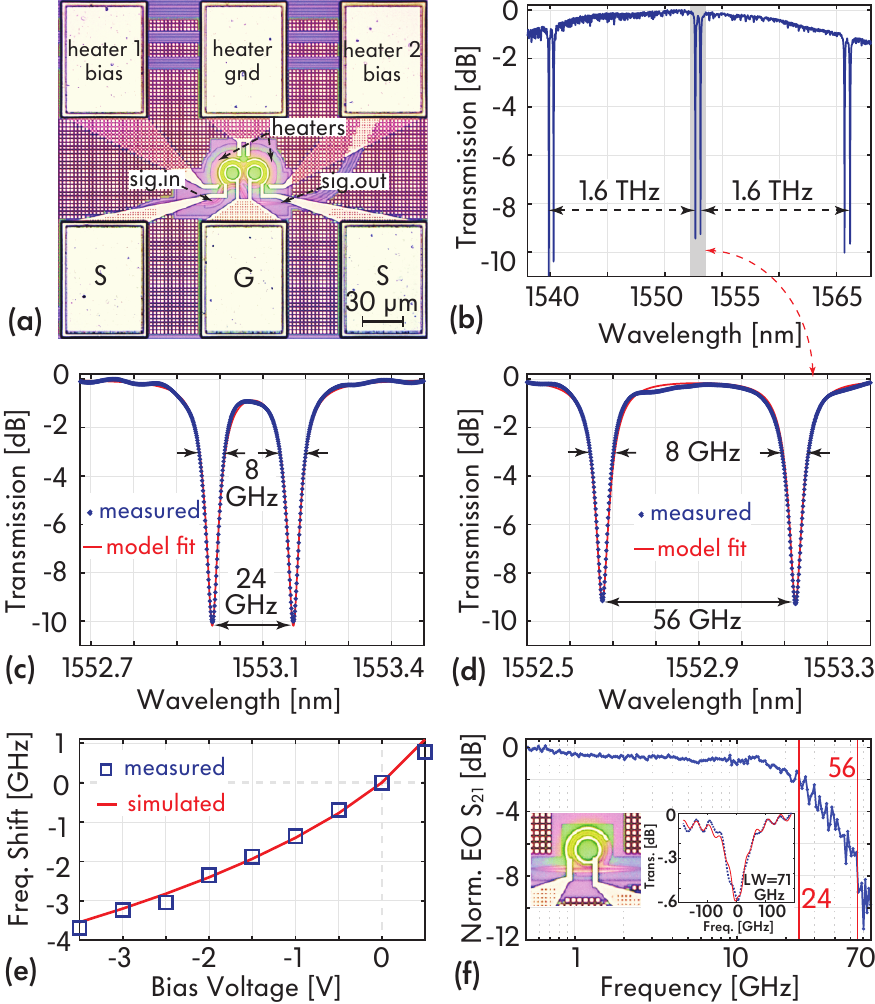}}
\caption{(a) Optical micrograph of the fabricated device. (b-d) Passive optical transmission spectra showing (b) multiple FSR resonances and (c,d) supermode resonances of the two device variants. (e) Cavity resonance frequency shift versus applied bias voltage. (f) Normalized RC-time-limited EO response of a strongly overcoupled (linewidth 71~GHz) microring modulator.}
\label{fig:dc_tests}
\end{figure}

The wavelength converter was fabricated in the IMEC ISIPP50G foundry silicon photonics platform \cite{Pantouvaki2017}. Figure~\ref{fig:dc_tests}(a) shows an optical microscope image of the device with metal pads wired to the EO phase shifters and microheaters. The passive optical transmission responses of the two device variants are shown in \mbox{Figs.~\ref{fig:dc_tests}(b-d)}. The devices have free spectral range (FSR) of 13~nm (1.6~THz) [\mbox{Fig. 2(b)}]. The total linewidths of the supermode resonances, shown in \mbox{Figs.~\ref{fig:dc_tests}(c,d)}, are similar across the two device variants and are equal to $\delta f_{tot}$=8~GHz, which corresponds to total quality factor of 24k, intrinsic Q of 70k and intrinsic linewidth of $\delta f_{int}$=2.7~GHz. The splitting in wavelength between the resonant supermodes is 0.19~nm (24~GHz) and 0.45~nm (56~GHz) for the devices with respective ring-to-ring coupling gaps of 445~nm and 315~nm. Resonator-embedded EO phase shifters produce 4.5~GHz shift in resonance frequency measured with 4V (-3.5 to 0.5~V) applied DC bias, as shown in \mbox{Fig.~\ref{fig:dc_tests}(e)}. To characterize phase shifter efficiency degradation at high electrical frequencies (RC bandwidth limit), we measured the small-signal EO $s_{21}$ response of a strongly overcoupled single-ring microring modulator with >70~GHz optical linewidth, which is fabricated on the same chip and uses an active cavity identical to the cavities in the wavelength converters [see inset in \mbox{Fig.~\ref{fig:dc_tests}(f)}]. The normalized response measured at 1.5~V reverse bias is shown in \mbox{Fig.~\ref{fig:dc_tests}(f)}. Due to the broad optical linewidth, the drop in the EO response at our frequencies of interest is primarily determined by RC time constant and is equal to 1.7~dB at 24~GHz, 3~dB at 30~GHz (3dB bandwidth), and 9~dB at 56~GHz. Assuming the wavelength converters are biased at -1.5~V and modulated with 4V peak-to-peak electrical drive and taking into account the 1.7 and 9~dB efficiency drop, from \mbox{Fig.~\ref{fig:dc_tests}(e)} we estimate the peak-to-peak resonance frequency swings $\delta f_{pp}$ to be 1.6 and 3.6~GHz at respective drive frequencies of 24 and 56~GHz. Substituting the values of $\delta f_{int}$, $\delta f_{tot}$, $\delta f_{pp}$, and $\Delta f = 0$ into Eq.~\ref{eqn:conveff} we estimate -11~dB peak conversion efficiency for the 24~GHz converter and -18~dB for 56~GHz converter.

\begin{figure}[!b]
\centering
\mbox{\includegraphics[width=\linewidth]{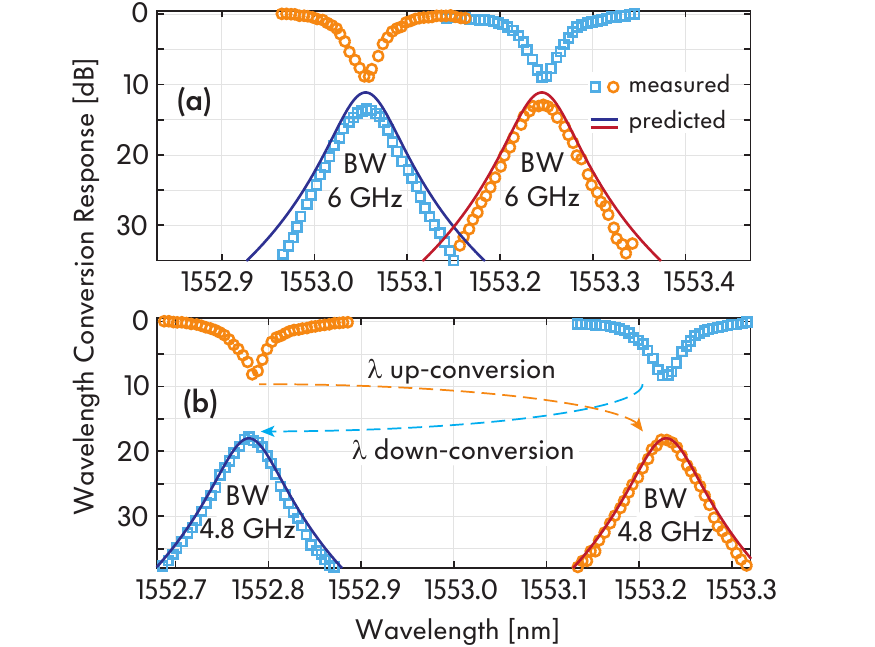}}
\caption{Wavelength up- and down-conversion responses of the two device designs normalized by the input optical power.}
\label{fig:conv_eff}
\end{figure}

Next we measured the dependence of the conversion efficiency on the wavelength of input light. The active cavities of the  wavelength converter are biased at -1.5~V and are driven by a 13~dBm differential microwave signal (10~dBm on each cavity) delivered to the converter through a 50~$\Omega$ RF cable and applied to the signal pads with an SGS probe. Taking into account the low-frequency voltage doubling the active cavities see about 4V peak-to-peak drive swing. CW laser light is coupled into and out of the chip through a pair of grating couplers. The wavelength of the CW light is stepped across each of the supermode resonances and for each wavelength setting the power of the converted signal and leftover signal at input wavelength is detected on an optical spectrum analyzer. The results for the two device designs are shown in Fig.~\ref{fig:conv_eff}. The data shown in orange (circles) and blue (squares) correspond to wavelength up- and down-conversion experiments, respectively. The dips in the wavelength spectrum represent the fraction of input power at input wavelength detected at the device output (leftover input), and the peaks represent the fraction of input power converted to a different wavelength. The converter designs performing 24 and 56~GHz shift show respectively -13 and -18~dB conversion efficiency and 6 and 4.8~GHz bandwidth. The solid lines in \mbox{Figs.~\ref{fig:conv_eff}(a,b)} show conversion efficiencies predicted by the theoretical model given by Eq.~\ref{eqn:conveff}. Note that the wavelength converter response in Fig.~\ref{fig:conv_eff} is analogous to the response of a second order add/drop filter, with symmetric and antisymmetric supermodes acting as ``resonators'' that comprise the filter and the push-pull modulation acting as ``coupling'' (now frequency translating) between them \cite{wade2014}. In the case of the wavelength converter, however, the ``through'' and ``drop'' responses appear at different wavelengths. We believe this to be the first demonstration of such a frequency translated filter response. The passband is more rounded than the usual flat-top shape of the filter because the loss is slightly dominating over the RF drive induced splitting. 

\begin{figure}[!t]
\centering
\mbox{\includegraphics[width=\linewidth]{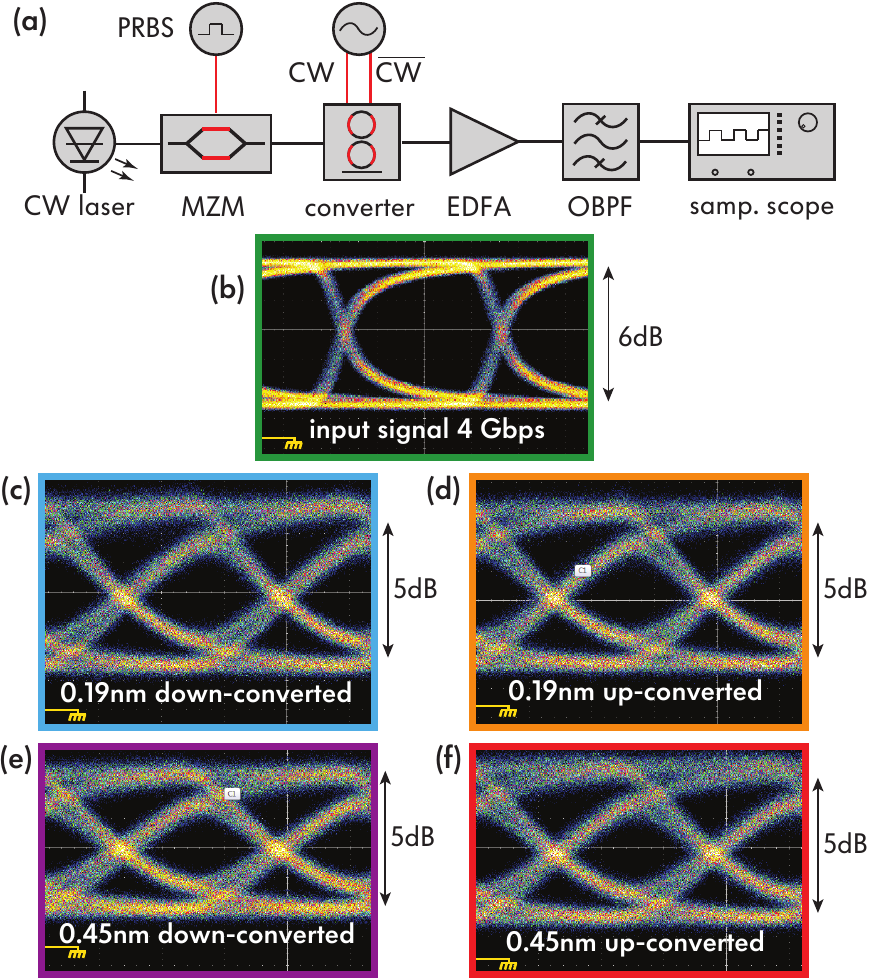}}
\caption{(a) Schematic of the experimental setup. (b) Optical eye diagrams of 4~Gbps input and (c-f) wavelength-shifted NRZ signals.}
\label{fig:eye_diag}
\end{figure}

We carried out wavelength conversion experiments on actual optical data streams. \mbox{Figure \ref{fig:eye_diag}(a)} shows a schematic of the setup used for this experiment. A CW laser, tuned to the wavelength of one of the supermode resonances, is modulated with a 4~Gbps NRZ pseudorandom bit sequence (PRBS) by a commercial lithium niobate modulator and is sent into the wavelength converter. The converter is again biased at -1.5~V and driven with a 4~V peak-to-peak drive signal at the corresponding 24 or 56~GHz frequency. The light at the converter output is amplified with an erbium-doped fiber amplifier (EDFA). The unwanted leftover input signal and the EDFA noise are suppressed by an off-chip optical bandpass filter (OBPF) with a 3.4~GHz-wide passband, tuned to the converted signal wavelength. Filtering of the signal with high extinction ratio can eventually be performed with high-order coupled microring resonator filters co-integrated with the converter on the same chip. Alternatively, a more sophisticated interferometric coupling carries out the filtering within the converter \cite{Gevorgyan2020}. Finally, the wavelength-converted NRZ signal is detected with a sampling scope. \mbox{Figure~\ref{fig:eye_diag}(b)} shows optical eye diagram of the input data stream, and \mbox{Figs.~\ref{fig:eye_diag}(c-f)} show eye diagrams of the wavelength up- and down-converted signals measured with both device variants. In this experiment, the data rate was limited to 4 Gbps primarily by the 3.4~GHz bandwidth of the external OBPF. Limited by the optical linewidth, however, these wavelength converters are expected to support data rates up to 8~Gbps. Larger bandwidths can be obtained using wider resonances at the expense of conversion efficiency.

In conclusion, the dual-active-cavity wavelength converter architecture demonstrated in this work is capable of converting optical signals to both higher and lower wavelengths with high conversion efficiency, requiring driving voltages that can be realized with CMOS circuits. This device concept implemented in CMOS electronic-photonic platforms and driven by high-frequency signal generated on the same chip would require no additional optical or RF input to the chip \cite{Stojanovic2018, Atabaki2018}, and could immediately become a new silicon photonics library component, opening the wavelength dimension in processing of signals. With the first demonstration of this device concept we achieved signal frequency shift of 24 and 56 GHz with respective efficiencies of -13 and -18~dB and bandwidths of 6 and 4.8~GHz and performed conversion of 4~Gbps NRZ signal. The conversion efficiency and bandwidth can be greatly improved by using more sensitive phase shifter designs \cite{watts2014} and by implementing an on-chip inductor that enhances the drive voltage \cite{Gevorgyan2020}. Such improvements will also allow wider bandwidths that support higher data rates at acceptable conversion efficiencies.

\section*{Funding Information}
Ball Aerospace and Technologies Corp. (IRAD FY2018); University of Colorado Boulder (2016 Innovative Seed Grant); Boston University (Innovation Career Development Professorship).

\section*{Acknowledgments}
We thank Joseph Bardin and Sayan Das, University of Massachusetts Amherst, for providing a high-frequency RF probe.

\section*{Disclosures}
A. Khilo: Ayar Labs, Inc. (E), M.A. Popovi\'{c}: Ayar Labs, Inc. (F,I,C)

% Bibliography
\bibliography{references}
\bibliographystyle{ieeetr}

\end{document}